# Colossal nonreciprocal Hall effect and broadband frequency mixing due to a room temperature nonlinear Hall effect

Lujin Min[1,2], Yang Zhang[3], Zhijian Xie[4], Sai Venkata Gayathri Ayyagari[2], Leixin Miao[2], Yugo Onishi[3], Seng Huat Lee[1,5], Yu Wang[1,5], Nasim Alem[2], Liang Fu[3*], Zhiqiang Mao[1,2*]

[1]Department of Physics, Pennsylvania State University, University Park, PA, USA

[2]Department of Materials Science and Engineering, Pennsylvania State University, University Park, PA, USA

[3]Department of Physics, Massachusetts Institute of Technology, Cambridge, MA, USA

[4]Department of Electrical and Computer Engineering, North Carolina Agriculture and Technical State University, Greensboro, NC, USA

[5]2D Crystal Consortium, Materials Research Institute, Pennsylvania State University, University Park, PA, USA

## Abstract

**Nonreciprocal charge transport in quantum materials has attracted enormous interest since it offers an avenue to investigate quantum symmetry related physics and holds many prospective applications such as rectification and photodetection over a wide range of frequencies.[1] The nonreciprocal transport reported to date occurs along the longitudinal direction with the nonreciprocal resistance limited to a few percent of the ohmic resistance.[2] Here we report a transverse nonreciprocal transport phenomenon with divergent nonreciprocity – colossal nonreciprocal Hall effect. This is revealed in direct current (DC) transport measurements on the microscale Hall devices made of the Pt wires deposited by focused ion beam (FIB) on Si substrates and the Weyl semimetal NbP with FIB-deposited Pt electrodes at zero magnetic field. When a DC is applied along the *x*-axis of the devices ($I_x$), it generates a voltage along the *y*-axis ($V_y$) near room temperature, with $V_y$ quadratically scaling with $I_x$. The transverse resistance, which shows a sign reversal upon switching the current direction, results from a colossal extrinsic nonlinear Hall effect (NLHE) rooted in the disorder scatterings in the Pt wires. While NbP was not found to show NLHE, the NLHE generated in the Pt electrodes can be transmitted to the NbP Hall devices, which yields a surprisingly large nonlinear anomalous Hall effect in NbP with the Hall angle ($\theta_H$) far exceeding the record value of the anomalous Hall angle of magnetic conductors at room temperature[3]. Furthermore, we find such a strong NLHE can lead to broadband frequency mixing, with the frequency spectrum of the Hall voltage including 2$^{nd}$-harmonic generation, sum & difference frequency generations, and other multiple (3-11) wave mixing components. These results not only demonstrate the concept of the nonreciprocal Hall effect for the first time but also pave the way for exploring NLHE's applications in terahertz communication, imaging, and energy harvesting.**

\* Corresponding authors. Email: liangfu@mit.edu (L.F.); zim1@psu.edu (Z.Q.M.)

Ohm's law is the most fundamental law governing electrical transport in materials and electronic circuits. It states that under a constant condition, the current flowing through a conductor is linearly proportional to the voltage across the two ends of the conductor when heating and junction effects are excluded. However, in non-centrosymmetric polar materials, there is a mechanism that can cause an experimentally observable deviation from Ohm's law under magnetic fields, i.e. the nonreciprocal transport.[2,4] When the applied current **I** and magnetic field **B** are perpendicular to each other and orthogonal to the polar axis **P** of a given polar material, as illustrated in Fig. 1a, the current-voltage (*I-V*) characteristic measured with a direct current (DC) can be described by $V = R_0 I + \gamma R_0 B I^2$ where $\gamma$ is a coefficient (Fig. 1b).[2] The corresponding resistance $R(I) \equiv \frac{V}{I} = R_0 + \gamma R_0 B I$ depends on the direction of the current, resulting in nonreciprocal transport. The degree of nonreciprocity is characterized by the ratio of the nonreciprocal resistance to the ohmic resistance: $\eta = \frac{R(I) - R(-I)}{R(I) + R(-I)}$. The value of $\eta$ is usually very small such that the violation of Ohm's law is hardly observable in DC transport measurements, and most nonreciprocal transport phenomena reported so far are instead probed by alternating current (AC) transport measurements (see Supplementary Table 1).[5-10]

Recently, a second-order transverse transport phenomenon induced by Berry curvature effect or disorder scattering, i.e. the so-called nonlinear Hall effect (NLHE), was proposed for noncentrosymmetric but time-reversal-invariant materials.[11,12] Considering a material with a polar axis **P**, when a DC $I_x$ is applied perpendicular to **P**, a transverse voltage $V_y$ proportional to the square of the current develops along **P**, as shown in Fig. 1c. The quadratic dependence $V_y \propto I_x^2$ (Fig. 1d) is a fundamental violation of Ohm's law. Importantly, upon reversing the direction of the applied current $I_x$ the transverse voltage is symmetric to $I_x$, so that the transverse resistance $R_{yx} \equiv \frac{V_y}{I_x} \propto I_x$ changes sign, thus leading to a divergent transverse nonreciprocity as defined by $\eta_H \equiv \frac{R_{yx}(I_x) - R_{yx}(-I_x)}{R_{yx}(I_x) + R_{yx}(-I_x)} \to \infty$. Such nonreciprocal transverse DC transport can be termed as a nonreciprocal Hall effect (NRHE). While the NLHE has been observed in several material systems through AC transport measurements[13-26], there have been no reports on DC *I-V* characteristics that demonstrate transverse nonreciprocity. Furthermore, the reported NLHEs are primarily low-temperature phenomena. Room-temperature NLHEs were seen only in exfoliated TaIrTe$_4$ flakes[15] and BaMnSb$_2$ bulk single crystals[22]. Nevertheless, even at the maximum applied current, the NLHE-induced transverse voltage in these materials is still orders of magnitude smaller than the longitudinal voltage at linear response caused by the unavoidable electrical contacts' misalignment, which prevents the observation of the NRHE and limits the potential NLHE-based applications.

In this work, we report the first observation of a room-temperature colossal NRHE characterized by divergent $\eta_H$. Such a NRHE arises from an exceptionally strong extrinsic NLHE, which is caused by disorder scattering in the Pt wire deposited by focused ion beam (FIB) on Si substrates. Surprisingly, we also found such a NLHE can be transmitted to a conductor when the FIB-deposited Pt (FIBD-Pt) wires are used as the electrodes to inject a current to the conductor, thus yielding not only a colossal NRHE but also an exceptionally large anomalous Hall angle $\Theta_H^A$ in the conductor at zero magnetic field. As an example, we show that the Weyl semimetal NbP, while not displaying an intrinsic NLHE, exhibits $\Theta_H^A$ of 39° at 1 mA and 330K in its microscale Hall devices with FIBD-Pt electrodes, which far exceeds the room temperature record value (12°) of $\Theta_H^A$ in magnetic conductors. Furthermore, we also demonstrated that the NLHE of the FIBD-Pt or the NLHE transmitted from the FIBD-Pt electrodes to NbP can be used for broadband frequency mixing and wireless microwave detection, which highlights the tremendous potential of using nonlinear Hall devices for terahertz (THz) communication, imaging, and energy harvesting.

This work was originally inspired by the earlier theoretical studies on the Weyl semimetal NbP [1,27] which predicted this material exhibits a strong intrinsic NLHE due to Berry curvature dipole (BCD). Therefore, it might serve as a platform for the demonstration of the NRHE. Under this motivation, we fabricated micrometer scale Hall devices of NbP single crystals through FIB machining (see Methods and Supplementary Note 1). The FIBD-Pt wires were used as electrodes and the devices were fabricated on silicon substrates. Using such a micro-scale Hall device, we successfully observed a colossal NRHE near room temperature. Nevertheless, we found such NRHE does not originate from NbP but is transmitted from the FIBD-Pt electrodes. Before we present the results obtained on the NbP devices, we first show evidence for the NRHE observed in the Hall device made of FIBD-Pt. Figure 1e shows a Hall device of FIBD-Pt (see Methods for the fabrication details). The DC I-V characteristics at various temperatures measured on this device (Pt-S1) with the current applied along the x-axis are shown in Fig. 1f. When the temperature is ≤ 250K, the I-V curves exhibit linear dependences, consistent with the expected Ohm's law behavior. In stark contrast, as the temperature is increased above 250K, the I-V curves start to significantly deviate from linear behavior and such a deviation reaches a maximum at temperatures close to 350K where the I-V curves are parabolic-like. Such a parabolic-like I-V curve becomes much more remarkable in a large current range (see Fig. 1g). By symmetrizing and anti-symmetrizing the measured data, the transverse voltage $V_y$ can be decomposed into a quadratic voltage component $V_y^{QD}(\propto I_x^2)$ (Fig. 1h) and a linear voltage component $V^L(\propto I_x)$ (see Supplementary Fig. S3a). $V^L$ is due to the electrical contacts' misalignment, while $V_y^{QD}$ corresponds to the second-order Hall voltage response as to be discussed below. Note that device Pt-S1 (Fig. 1e) was prepared in two steps: we first made a Pt cross with each arm ~10 μm and then extended each arm to ~100 μm. We did this to mimic the NbP device structure (inset to Fig. 2b). This was indeed unnecessary. We reproduced the parabolic DC I-V characteristic in both the cross devices formed by two continuously deposited Pt wires and the FIBD-Pt Hall bar device which used evaporated Au pads as electrodes (see Supplementary Note 2 & Fig. S4).

The parabolic I-V characteristics near 350 K (Fig. 1f-1h) clearly demonstrate the sign reversal of the transverse Hall resistance between negative and positive currents. From the raw I-V curve at 350 K of device Pt-S1, the nonreciprocity coefficient $\eta_H$ is estimated to be 349% for $I_x$ = 50 μA and 2068% for 500 μA (Fig. 1g). After eliminating the $V^L$ component via symmetrizing the data, the transverse voltage shows exact quadratic dependence on the applied current ($V_y^{QD} \propto I_x^2$) (Fig. 1h), so that $\eta_H$ is infinitely large. All these features agree well with the expected NRHE. We also performed DC transport measurements with $I \parallel y$ and found the transverse nonlinear voltage response along the x-axis is smaller than that for $I \parallel y$, which is due to the difference in disorder scattering between the x and y axes as explained in Supplementary Note 2.

Through AC transport measurements using the lock-in technique, we have further confirmed that the NRHE observed in the FIBD-Pt devices originates from the NLHE. As shown in Fig. 1i, when an AC was applied along the x-axis of device Pt-S1 within the temperature window where NRHE was observed, a transverse second-harmonic voltage along the y-axis, $V_y^{2\omega}$, was detected and $V_y^{2\omega}$ scales quadratically with the input current $I^\omega$ and remains independent of frequency below 1000 Hz (see the inset to Fig. 1i). Moreover, a rectified Hall voltage $V_y^{RDC}$ along the y-axis, which is about 1.4 times $V_y^{2\omega}$ and proportional to $(I^\omega)^2$, was also detected. $V_y^{RDC}$ is almost equal to the $V_y^{QD}$ component probed in the DC transport measurements. $V_y^{2\omega}$, $V_y^{RDC}$ and $V_y^{QD}$ all exhibit the same temperature dependence and show maxima at 350K (Fig. 1j). These observations indicate that the FIBD-Pt generates a strong NLHE, which is responsible for the observed NRHE. This is further verified in the measurements on the FIBD-Pt Hall bar device with Au electrodes (see Supplementary Note 2). Although the second harmonic voltage response can also be driven by some other mechanisms such as the junction effect and the thermoelectric effect, our observations of frequency independence of $V_y^{2\omega}$ and quadratic DC I-V characteristics exclude those mechanisms (see Supplementary Note 3 for detailed discussions).

As noted above, the NLHE generated in the FIBD-Pt can be transmitted to a conductor when the FIBD-Pt wire is used as electrodes to inject a current into the conductor. This is revealed in our measurements on the microscale NbP devices with FIBD-Pt electrodes. The inset to Fig. 2b shows an image of a representative cross-shape Hall device made of NbP and FIBD-Pt electrodes (NbP-S1). In DC transport measurements on this device, we observed a NRHE similar to that seen in device Pt-S1. Fig. 2a presents the I-V characteristics of NbP-S1 at various temperatures. Like device Pt-S1, the NbP-S1 device also exhibits parabolic-like I-V curves at temperatures near 350 K and this feature also becomes much more significant as the current range is increased to the -0.5-0.5 mA range (Fig. 2c). Through symmetrizing the measured Hall voltage, we also obtained a quadratic Hall voltage component $V_y^{QD}$ ($\propto I_x^2$, see Fig. 2b) comparable to that seen in Pt-S1 for the same applied current, e.g. $V_y^{QD}$ = 0.81 mV for NbP-S1 and 0.75 mV for Pt-S1 at $I_x$ =50 μA and $T$=350 K (see Fig. 1h & 2b). These data demonstrate that a colossal room-temperature NRHE with divergent $\eta_H$ is also present in NbP. This result was also reproduced in the other cross-like NbP device (NbP-S2, see Supplementary Note 4 & Fig. S6). However, as stated above, such a NRHE is not generated by NbP, but transmitted from the FIBD-Pt electrodes attached on NbP. This is revealed from the following experiments. First, in NbP-S1, we also probed frequency-independent second-harmonic Hall voltage $V_y^{2\omega}$ below 2000Hz and rectified Hall voltage $V_y^{RDC}$ which quadratically scales with $I^\omega$ (Fig. 2d); both $V_y^{2\omega}$ and $V_y^{RDC}$ exhibit the nearly same temperature dependence as those seen in device Pt-S1, with their maxima all appearing near 350 K, as shown in Fig. 1j. This similarity suggests that the NLHE generated in FIBD-Pt is transmitted to NbP.

Second, to further demonstrate the nonlinear Hall voltage response probed on devices NbP-S1 &S2 is due to NLHE, but not from other origins, we also fabricated a standard NbP Hall device with two pairs of FIBD-Pt Hall voltage leads (NbP-HB1) (Fig. 2e). Using such a device, we proved a nonlinear Hall angle $\theta_{\mathrm{NLHE}}$ close to 90° in the conductor (NbP) with the NLHE transmitted from the FIBD-Pt electrode. $\theta_{\mathrm{NLHE}}$ is defined as $\arctan(V_y^{2\omega}/V_x^{2\omega})$ where $V_x^{2\omega}$ represents the longitudinal voltage at the frequency of 2ω. Considering the second-order response is generated by FIBD-Pt instead of NbP, the 2$^{nd}$-order voltage response occurs only along the transverse direction, but not along the longitudinal direction (i.e. $V_x^{2\omega}$=0), thus $\theta_{NLHE}$ = 90°. However, in reality, there are always unavoidable misalignments of the voltage test leads that can lead the longitudinal voltage to involve a small Hall voltage component, thus engendering the deviation of $\theta_{NLHE}$ from 90°. This is exactly what we observed in device NbP-HB1. Fig. 2f shows the second-harmonic voltages measured on four different pairs of voltage leads for NbP-HB1 at 330K where its $V_y^{2\omega}$ is maximal, i.e. $V_{12}^{2\omega}$, $V_{34}^{2\omega}$, $V_{24}^{2\omega}$, and $V_{13}^{2\omega}$. Here the subscript numbers (1, 2, 3 & 4) represent the Hall voltage leads shown in Fig. 2e. All these measured second-harmonic voltages quadratically depend on the driving current (Fig, 2f). The transverse voltages $V_{12}^{2\omega}$ and $V_{34}^{2\omega}$ are much larger than the longitudinal voltages $V_{13}^{2\omega}$ and $V_{24}^{2\omega}$. The extracted $\theta_{NLHE}$ is ~72-84°, consistent with the $\theta_{NLHE}$ = 90° expectation. Such nonlinear Hall voltage's dominance over the longitudinal response, together with the frequency independence of $V_y^{2\omega}$ and the linear DC *I-V* characteristics measured using the two-probe method (see Supplementary Fig. S5) exclude the junction effect from the contacts or thermoelectric effect. DC measurements on Nb-HB1 also reproduce NRHE as seen in NbP-S1 (Supplementary Fig. S7). In addition, we proved that NbP does not exhibit intrinsic NLHE. We prepared another type of microscale NbP cross-shape device via polishing and FIB machining. These devices did not use FIBD-Pt as electrodes; instead, the Au wires glued on the sample by silver epoxy were used as electrodes (see Supplementary Note 5 and Fig. S9). As shown in Supplementary Fig. S9, we did not observe any 2$^{nd}$-harmonic Hall voltage response along the polar axis of NbP (i.e. the *c*-axis) even when the current density applied along its *a*-axis is comparable to that applied on the NbP devices with FIBD-Pt electrodes (NbP-S1 & NbP-HB1, Fig. 2b & 2e). Given intrinsic NLHE was theoretically predicted, its absence in our NbP devices with Au electrodes is most likely due to its Fermi level far away from the nodal rings which are expected to generate BCD-induced NLHE[1,27].

The NLHE transmitted from FIBD-Pt to NbP generates an exceptionally large anomalous Hall angle $\theta_H$ in NbP. This is revealed in the DC transport measurements on the NbP-HB1 device (Fig. 2e). In those measurements, the DC is applied along the long-dimension direction of the device (i.e. the *x*-axis) while the Hall voltage is measured along the *y*-axis. The measurements were performed under zero magnetic field. In general, $\theta_H$ is defined as $\tan^{-1}(\frac{E_y}{E_x})$ (see the inset to Fig. 3a), where $E_x$ and $E_y$ represent the longitudinal and transverse (Hall) field respectively. Since the Hall field probed in our experiment is generated by the NLHE, the Hall field $E_y$ should be proportional to the square of $E_x$. Hence, $\tan\theta_H \propto E_x$. In other words, $\tan\theta_H$ should be linearly dependent on the applied current $I_x$. This is exactly what we observed in device NbP-HB1. Figure 3a presents the $\tan\theta_H$ of NbP-HB1 as a function of $I_x$ measured at 330K where its NLHE is close to the maximum (Supplementary Fig. S7). The value of $\tan\theta_H$ reaches 0.81 at 1 mA, which corresponds to $\theta_H$ = 39°. We also estimated the $\tan\theta_H$ value in the FIBD-Pt devices, which reached 0.89 at 0.1mA. To the best of our knowledge, this angle is much larger than any of the previously reported anomalous Hall angles of magnetic conductors[3,28-35] (Fig. 3b) and can be further increased with a larger current. Nonlinear Hall materials with large Hall angles are recently predicted to be promising candidates for quantum rectification and can find applications in wireless charging and energy harvesting.[36]

In the longitudinal transport measurements on the NbP-HB1 device, we also found that when the applied current is large enough, the longitudinal voltages deviate from linear dependence on the input current, as shown in Fig. 3d which plots the $I_x$-$V_x$ data measured at 330 K. As $I_x$ is increased above 0.4 mA, $V_x$ displays sublinear dependence on $I_x$, which becomes more pronounced as $I_x$ approaches 1 mA. Such a sublinear $I_x$-$V_x$ curve can be well fitted by including the third-order correction, as shown by the blue fitted curve in Fig. 3d. Furthermore, we also found that when the linear relation between $I_x$ and $V_x$ breaks down, $I_x$-$V_y$ relation deviates from the quadratic relation. Figure 3c presents the $I_x$-$V_y$ data measured in NbP-HB1 at 330K, from which we can see a clear deviation of $V_y$ from being quadratic in $I_x$ as $I_x$ approaches 1 mA. Such $I_x$-$V_y$ data can be fitted by including the fourth-order correction, as shown by the blue fitted curve. Such a deviation of $V_y$ from the quadratic dependence on $I_x$ in the large current range was also observed in Pt-S1 (Fig. 1g) and NbP-S1 (Fig. 2c). Both the third-order correction of $I_x$-$V_x$ and the fourth-order correction of $I_x$-$V_y$ can be attributed to high-order nonlinear response, as predicted in a recent theory [36] (see Supplementary Note 6 for discussions). Moreover, we also observed an anomaly in the longitudinal resistivity near the temperature where the nonlinear Hall response is maximal, which further demonstrates the coupling between longitudinal and transverse transport (see Supplementary Note 6 & 7 for details).

As noted above, the NLHE generated in FIBD-Pt is of extrinsic origin and caused by disorder scattering. Previous studies have shown that the FIBD-Pt is an amorphous mixture of Pt nanoparticles dispersed in Ga and C, devoid of any crystalline symmetry[37]; this is also verified in our composition and structure analyses for the FIBD-Pt in our devices (see Supplementary Note 8). Hence, the NLHE effect observed in the FIBD-Pt cannot stem from the Berry curvature dipole which is present only in noncentrosymmetric crystalline material with non-trivial band topology. However, from a macroscopic perspective, the inversion symmetry of FIBD-Pt can still be locally broken by the texture of Pt nanoparticles formed in the FIB scanning path (see Supplementary Fig. S2). Pt, the main component contributing to the conductivity of this mixture, is renowned for its strong spin-orbital coupling (SOC). Moreover, the FIBD-Pt features rich grain boundaries and a high concentration of disorders. The combination of these characteristics meets the prerequisites for creating extrinsic NHLE via skew and/or side jump scattering[12,14,19]. Additionally, our experiments also found that the NLHE of the FIBD-Pt depends on the substrate used for device fabrication. NLHE was observed only in the Pt deposited on bare Si substrates via FIB, but not in the Pt deposited on the Si wafer coated with a $SiO_2$ layer (see Supplementary Note 9 and Fig. S15). This suggests that the Si/FIBD-Pt interface plays a critical role in creating asymmetric scattering, which is necessary for generating extrinsic NLHE. This is further evidenced by the observation that the decrease in the thickness of the FIBD-Pt wire leads to an enhanced NLHE (see Supplementary Note 10). Our scaling analysis of the nonlinear Hall responsivity $E_y/(E_x)^2$ vs longitudinal conductivity $\sigma_{xx}$ does not reveal signatures of skew scattering/side jump dominated NLHE (Supplementary Note 11). This can possibly be attributed to the fact that the amorphous mixture of Pt, Ga, and C involves strong boundary scattering, which significantly increases resistivity, but not strikingly contributes to asymmetric scattering, thus leading to the breakdown of the scaling law of skew scattering/side jump (see supplementary 11 for detailed discussions). Besides the NLHE observed in FIBD-Pt, we also found that the W wires deposited on Si substrate via FIB also exhibit NLHE (see Supplementary Note 9). This is not surprising since W is also a heavy element and should also bring in strong SOC, thus enabling skew and/or side jump scattering.

Materials with strong NLHE have been predicted to be useful in many practical applications such as THz communication/imaging, and energy harvesting. Of course, real applications require strong NLHE at room temperature. The NLHE of the FIBD-Pt appears to meet this requirement, as manifested in its large room temperature nonlinear Hall responsivity $E_y/(E_x)^2$ (see Supplementary Table 2). In this work, we discovered that the large NLHE of the FIBD-Pt can enable another new functionality – broadband electronic frequency mixing. As shown in the inset of Fig. 4a, two voltage sources with identical output voltages but different frequencies were connected in parallel to two opposite ends of a cross-shaped nonlinear Hall device (NbP-S1, inset to Fig. 2b). A lock-in amplifier was used to measure the transverse output Hall voltage. When two alternating voltages of 125Hz and 37Hz were applied to the device, we observed 21 distinct frequency peaks, ranging from 20Hz to 280Hz. This frequency spectrum includes the first harmonic signal due to the contact misalignments, the second harmonic generation (SHG; $2\omega_1$ & $2\omega_2$), the sum frequency generation (SFG, $\omega_1+\omega_2$), difference frequency generation (DFG, $\omega_1-\omega_2$), and 17 other multiple wave mixing (WM) components. Among those harmonic peaks, $V^{\omega_1+\omega_2}$ and $V^{\omega_1-\omega_2}$ are the strongest. All other signal pairs at conjugacy frequencies, i.e., $a\omega_1 \pm b\omega_2$, exhibit equal amplitude. Notably, the eleven-wave-mixing component (11WM, $4\omega_1 - 6\omega_2$) at 278Hz can still be observed. Here it should be mentioned that although we labeled the mixed frequency peaks with the lowest-order effects, they also have contributions from the higher-order nonlinear process. For instance, $V^{\omega_1-\omega_2}$ and $V^{\omega_1+\omega_2}$ are more than two times larger than $V^{2\omega_1}$ or $V^{2\omega_2}$, inconsistent with the theoretical expectation of $V^{\omega_1-\omega_2}$ (or $V^{\omega_1+\omega_2}$) = 2 $V^{2\omega_1}$ or $2V^{2\omega_2}$ (see Supplementary Note 12). Such unusually large $V^{\omega_1-\omega_2}$ and $V^{\omega_1+\omega_2}$ can only be attributed to higher-order contributions.

The NLHE origin of such frequency mixing was further verified by examining their scaling with the input voltage amplitude. We maintained one input voltage, $V_{in}^{\omega_1}$, as a constant and performed $V_{in}^{\omega_2}$ dependent measurements on the output Hall voltage $V_{out}$ with the lock-in amplifier locked at several characteristic mixed frequencies. As the input voltage $V_{in}^{\omega_2}$ is small, high-order contributions can be negligibly small. For the SFG and DFG, $V_{out}$ is linearly proportional to the input voltage $V_{in}^{\omega_2}$ (Fig. 4b). When the mixed frequencies involve the $2\omega_2$ component, their Hall voltage signals quadratically scale with $V_{in}^{\omega_2}$ (Fig. 4c). When the $3\omega_2$ component is involved, e.g., the 7WM signal of $3\omega_1 - 3\omega_2$, $V_{out}$ scales cubically with $V_{in}^{\omega_2}$ (Fig. 4d). We reproduced a similar frequency mixing functionality in the Pt-S1 device and showed such frequency mixing is not due to junction effects (see Supplementary Note 13).

Further, we also demonstrated such frequency mixing can be extended to GHz ranges. As shown in Fig. 4e, when the input signal frequencies are 1 GHz and 1.04 GHz respectively, we observed two SHG signals at 2 GHz and 2.08 GHz and one SFG signal at 2.04 GHz. Given that signals up to the tenth order can still be observed near the DC limit, there is a possibility that the strong NLHE of the FIBD-Pt can be utilized to develop a new type of THz source. If this is proven to be true, it may find applications in THz imaging and communication. Compared to traditional frequency mixers based on nonlinear circuits, the NLHE-based frequency mixer has a much simpler structure with only a single material. Additionally, the NLHE mixer can spontaneously separate even-order from odd-order signals when electrical contacts' misalignment is eliminated.

In addition to frequency mixing, we have also demonstrated the nonlinear Hall devices made of either FIBD-Pt or NbP with FIBD-Pt electrodes can function as sensors to wirelessly detect microwave frequencies. As shown in Fig. 4f, a signal generator generated a 2.4GHz radio frequency (RF) signal which was transmitted to the devices through a pair of antennas. Meanwhile, a signal analyzer, connected to the other pair of terminals in the transverse direction, detected a strong second-harmonic signal. The power of the detected signal exhibits a quadratic relationship to the output power of the source, as shown in Fig. 4g, aligning well with the NLHE properties revealed in transport measurements.

## Methods
### Crystal Growth
The single crystals of NbP were synthesized using a chemical vapor transport (CVT) method via the following procedures.[38] Firstly, the stoichiometric mixture of Ta and P powder was sealed in an evacuated quartz tube and sintered at 800°C for 48 hours. Secondly, we ground the sintered NbP polycrystalline material into powder inside a glovebox and then sealed it into another evacuated quartz tube with iodine as a transport agent (13.5 mg cm$^{-3}$). Then the quartz tube was placed into a double-zone furnace with the temperature set at 850°C for the cold zone (source) and 950°C for the hot zone (sink). After one week of the CVT growth, large shiny NbP single crystals with flat (0 0 1) surfaces can be found near the hot end. We confirmed the grown NbP crystals have the desired structure and composition from X-ray diffraction and Energy-dispersive X-ray spectroscopy analyses.

### Micrometer Hall Device Fabrication
Microscale NbP devices in the cross and Hall bar geometry were fabricated by focused ion beam (FIB) cutting using FEI Helios NanoLab 660 and FEI Scios 2 dual beam SEM. First, we cut NbP thin lamellar samples with typical dimensions of 15 × 15 × 3μm (cross devices) or 20 × 10 × 2μm (Hall bar device) from millimeter-scale crystals using FIB and then lift and transferred it to Si substrates. The FIB cutting was made along the *a*-axis on the (0 0 1) surface so that the edges of the lamellar sample are approximately parallel to the crystallographic *a*- and *c*-axis respectively. Second, we further cut the lamellar crystal into a cross-shape or bar-shape with FIB and deposited Pt wires and pads for electrical contacts (see Supplementary Note 1 for detailed procedures). Microscale Pt devices were directly deposited to Si substrate. FIBD-Pt devices showing NLHE were fabricated by depositing Pt on Si substrates using FIB (see supplementary Fig. S2a). FIBD-Pt devices fabricated on the Si substrates coated with SiO$_2$ do not show NLHE (see supplementary Fig. S15b). In our previous report on the NLHE of BaMnSb$_2$ (ref. 22), we used Si substrates coated with SiO$_2$ for its device fabrication, and all the BaMnSb$_2$ crystals were glued on the substrate using GE-varnish. Both SiO$_2$ and GE-varnish should prevent the presence of extrinsic NLHE from FIBD-Pt.

**Electrical Transport Measurements**

The electrical transport measurements were conducted inside a Quantum Design Physical Property Measurement System (PPMS). The conventional linear Hall measurement data are shown in Supplementary Note 14. As for the nonlinear measurements, the AC and DC driving currents were, generated by a Keithley 6221 and a Keithley 6220 precision current source, respectively. The AC and DC voltages were measured by a Stanford Research SR860 lock-in amplifier and a Keithley 2182A nanovoltmeter, respectively.

The frequency mixing measurement was carried out using two Stanford Research SR860 lock-in amplifiers, connected in parallel as voltage sources, and a Stanford Research SR830 lock-in amplifier serving as the detector.

**Radio Frequency Measurement**

RF signals for both the frequency mixing experiment and the wireless detection measurement were generated by Keysight MXG Vector signal generators and were analyzed using an Agilent CXA signal Analyzer. In the frequency mixing experiment, the signal generators and analyzer were directly interfaced with the device. For the wireless detection measurement, wireless signals were transmitted and received using a pair of TP-link antennas. The receiving antenna was attached to one arm of the cross-like device, while the opposite arm was grounded.

**Data availability**

The authors declare that all the data that support the findings of this study are available within the paper and Supplementary Information. Additional relevant data are available from the corresponding authors upon reasonable request. Source data are provided with this paper.

## Acknowledgments

This work is primarily supported by the US National Science Foundation (NSF) under grant DMR 2211327. L.Min, L.Mia, N.A., S.V.G.A., N.A., and Z.M. also acknowledge the partial support from NSF through the Materials Research Science and Engineering Center DMR 2011839 (2020 - 2026). S.H.L. & Y.W. acknowledge the partial support by NSF through the Penn State 2D Crystal Consortium - Materials Innovation Platform (2DCC-MIP) under NSF Cooperative Agreement No. DMR-2039351. The work at Massachusetts Institute of Technology was supported by the U.S. Army Research Laboratory and the U.S. Army Research Office through the Institute for Soldier Nanotechnologies, under Collaborative Agreement Number W911NF-18-2-0048. L.F. was partly supported by the David and Lucile Packard Foundation. Y.O. thanks for the support from Funai Overseas Scholarship.

## Author contributions

The crystal growth and transport measurements were carried out and analyzed by L.Min and Z.M. The device fabrications were performed by L.Min, L.Mia, S.H.L., Y.W., N.A., and Z.M. The TEM study was conducted by S.V.G.A. and N.A. Theoretical support was provided by Y.Z., Y.O., and L.F. The idea of frequency mixing via NLHE was proposed by L.F. Z.X. performed the GHz frequency mixing and wireless microwave detection experiment. The paper was written by L.Min, Y.O., L.F., and Z.M. with inputs from other authors. N.A. and Z.M. supervised the experimental part of this work and L.F. supervised the theoretical part.

## Competing interest declaration.

The authors declare no competing interests.

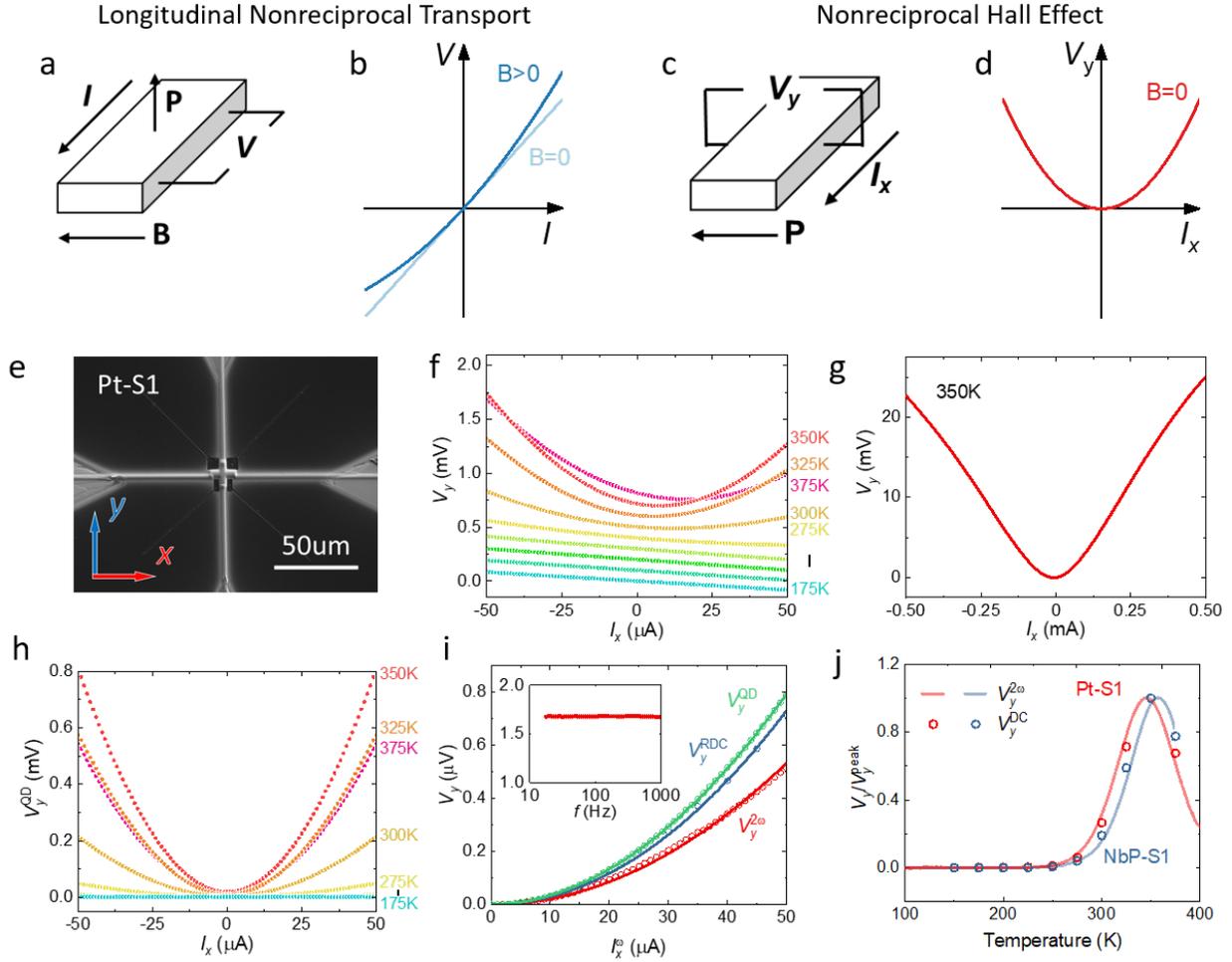

**Fig.1 Nonreciprocal Hall effect in FIBD-Pt.** (**a**&**c**) Schematic of the longitudinal nonreciprocal (NR) transport (**a**) and the nonreciprocal Hall effect (**c**) in a polar system with a polar axis along the *z*-axis. (**b** & **d**) Schematic DC *I-V* curves of the Ohmic transport (light blue in **b**), the longitudinal nonreciprocal transport (dark blue in **b**), and the nonreciprocal Hall effect (**d**). (**e**) SEM image of a cross-shape Hall device of FIBD-Pt (labeled with Pt-S1). (**f**) DC *I-V* characteristics measured with the current applied along the *x*-direction at different temperatures. (**g**) DC *I-V* characteristics measured in a larger current range at 350K. (**h**) The nonreciprocal Hall voltage component $V_y^{QD}$ vs $I_x$ at different temperature. (**i**) The second-harmonic Hall voltage $V_y^{2\omega}$, rectified Hall voltage $V_y^{RDC}$, symmetrized DC hall voltage $V_y^{QD}$ vs. the input current $I^\omega$ at 350K for Pt-S1. Inset: second-harmonic voltage $V_y^{2\omega}$ as a function of the input current frequency. (**j**) Normalized second-harmonic and rectified Hall voltages ($V_y^{2\omega}$, $V_y^{RDC}$) as a function of temperature for Pt-S1 and NbP-S1. $V_y^{2\omega}$ and $V_y^{RDC}$ are presented with lines and filled circles respectively.

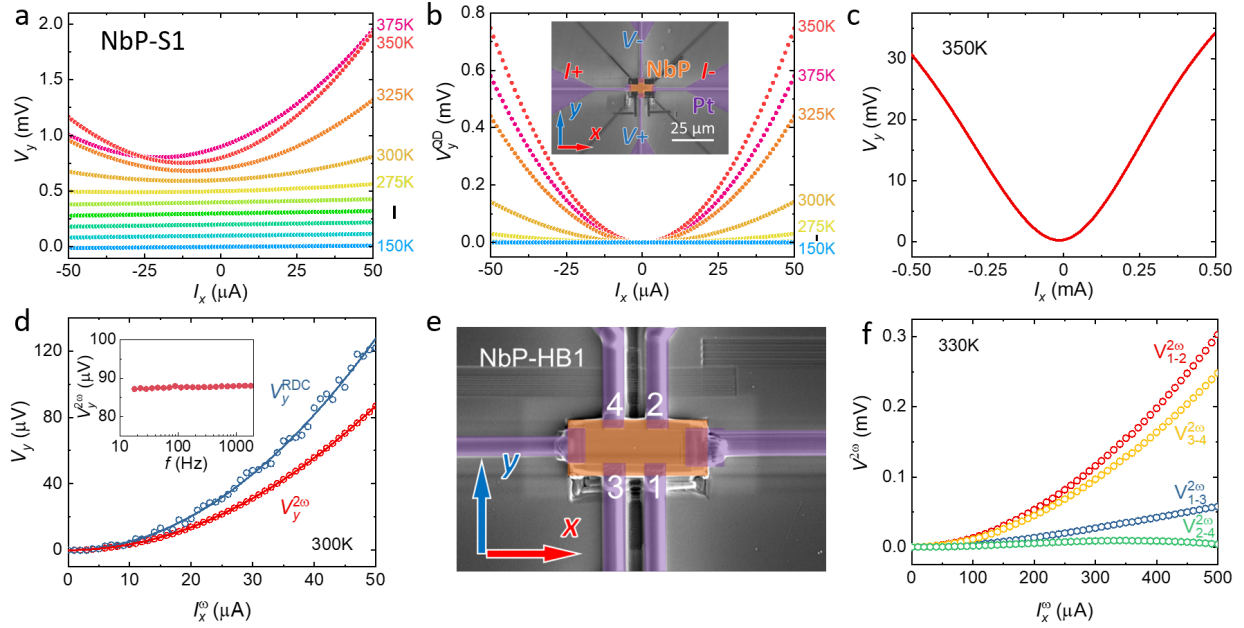

**Fig.2 Nonreciprocal Hall effect transferred to NbP**. (**a**) DC *I-V* characteristics measured with the current applied along the *x*-direction at different temperatures. (**b**) The nonreciprocal Hall voltage component $V_y^{QD}$ vs $I_x$ at different temperature. Inset: SEM image of NbP-S1. (**c**) DC *I-V* characteristics measured in a larger current range at 350K. (**d**) The second-harmonic Hall voltage $V_y^{2\omega}$ and rectified Hall voltage $V_y^{RDC}$ vs the input current $I^\omega$ at 300K for NbP-S1. (**e**) SEM image of NbP-HB1. The NbP crystal is colored in orange, while the Pt wires are colored in purple. (**f**) The second-harmonic Hall voltages as functions of input current measured between different pairs of contacts on NbP-HB1.

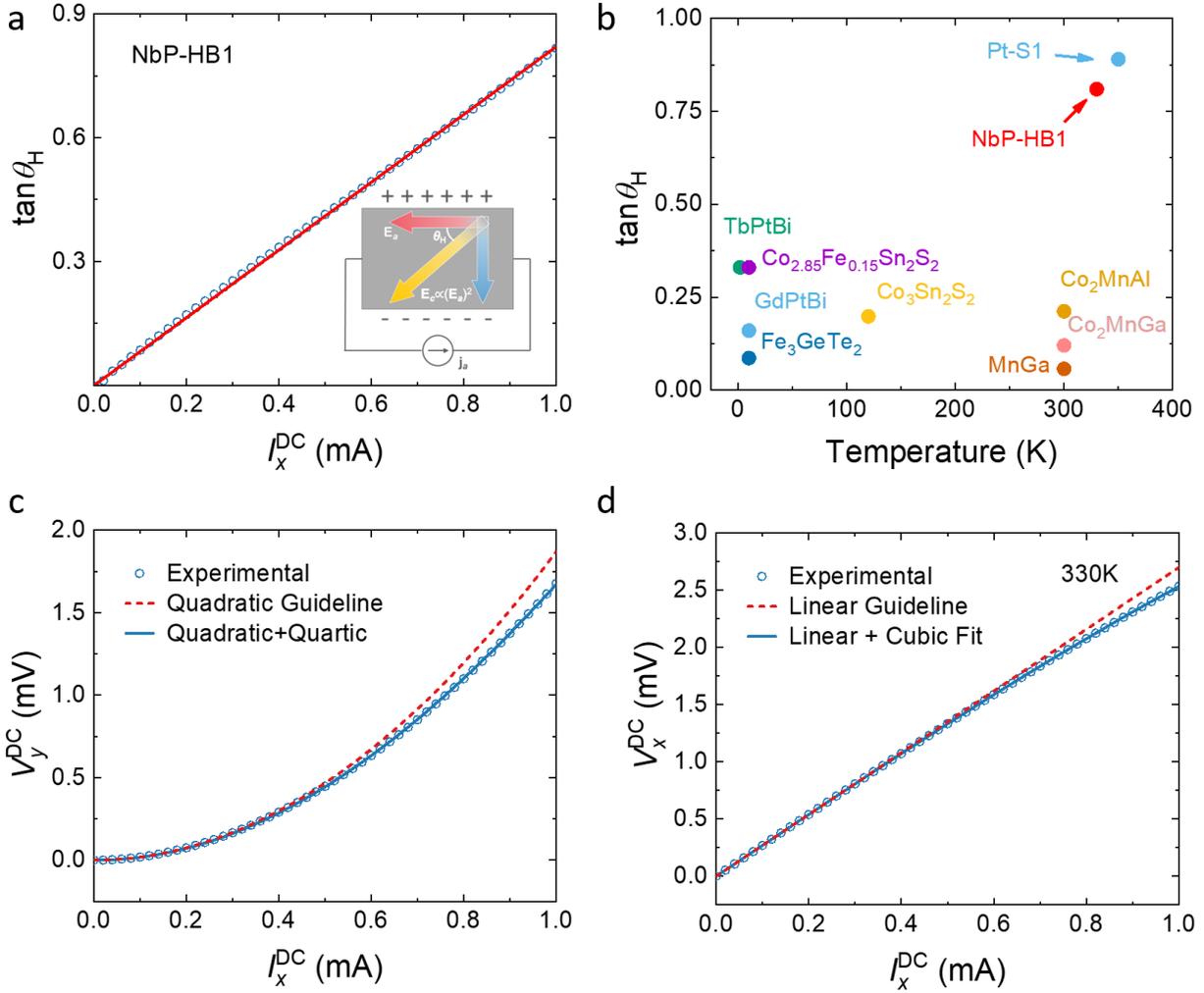

**Fig.3 Large nonlinear anomalous Hall effect in NbP transferred from Pt.** (**a**) Nonlinear anomalous Hall angle $\tan\theta_H$ vs input current $I_x^{DC}$. The linear fitted line is shown in red. The Inset: schematic of the electric field distribution in NLH material. (**b**) Comparison between the nonlinear anomalous Hall angle to the other well-known ferromagnetic materials. Both Pt and NbP samples exhibit much larger anomalous Hall angles compared with other materials at room temperature. (**c&d**) Transverse (**c**) and longitudinal (**d**) DC *I-V* characteristics of NbP-HB1. At high current range, the transverse/longitudinal experimental data start to deviate from the quadratic/linear *I-V* relationship. The discrepancies can be respectively corrected by an additional quartic/cubic term. The experimental data, one-component fitting, and two-component fitting are represented by blue hollow circles, red dashed lines, and blue solid lines, respectively.

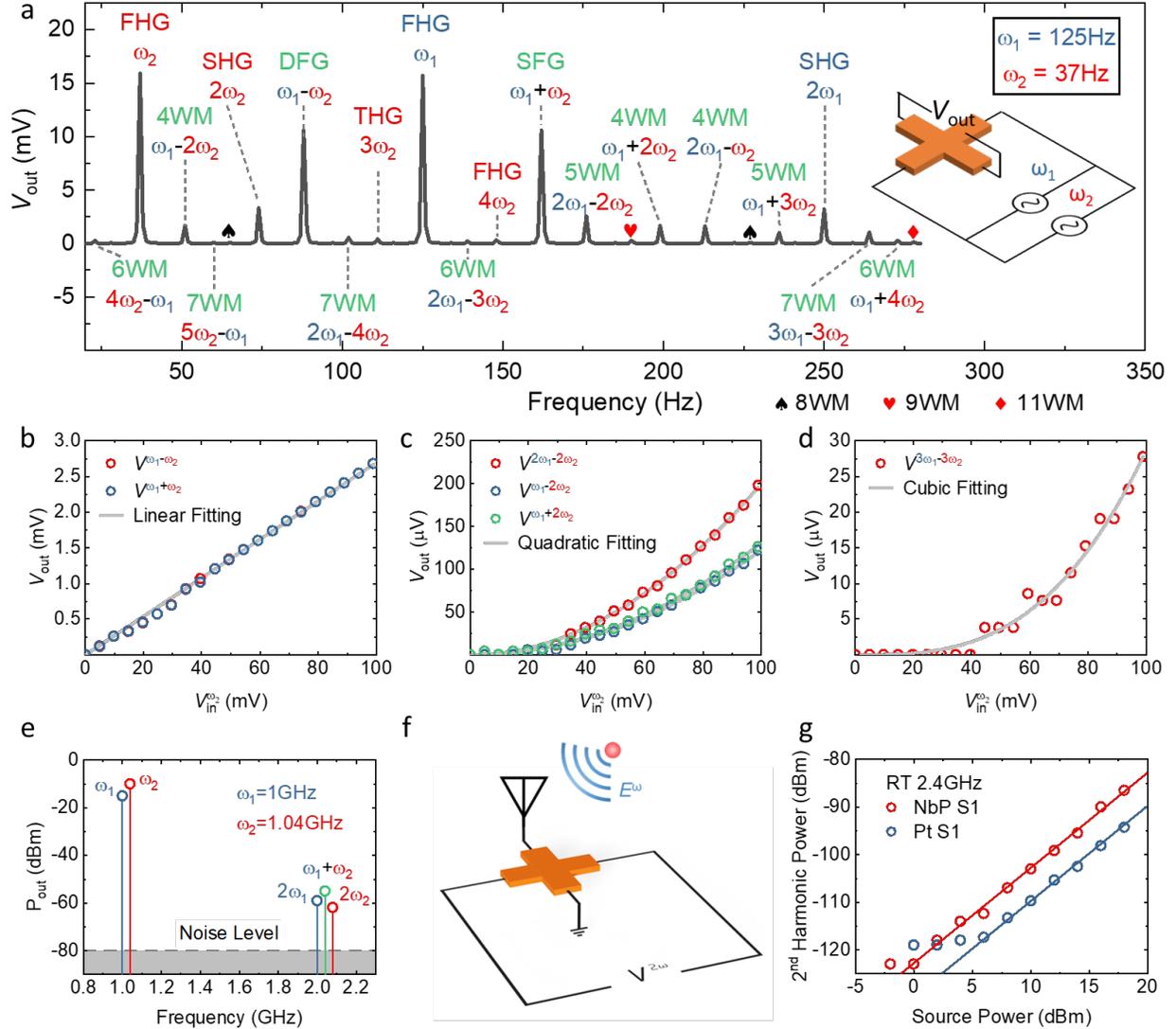

**Fig. 4 Broadband frequency mixing and microwave detection based on NLHE.** (**a**) Mixed frequency spectrum generated when two AC voltages of the same amplitude and different frequencies ( $\omega_1$ = 125Hz, $\omega_2$ = 37Hz) are simultaneously applied to device NbP-S1 at 350K. Mixed signals up to the 10th order (11WM) are still discernable. Blue and red labels denote the harmonic signals generated solely by voltages of frequencies $\omega_1$ and $\omega_2$ respectively. Green labels indicate mixed frequencies involving both input voltages. "NWM" stands for "N-wave-mixing". Inset: schematic of the experiment setup. (**b,c,d**) Dependence of the (**b**) SFG, DFG, (**c**) 4WM, 5WM, and (**d**) 7WM output voltages $V_{out}$ on $V_{in}^{\omega_2}$. The experimental data and fitted curves are shown in hollow circles and solid lines respectively. (e) Room temperature frequency mixing spectrum generated when two AC signals with frequencies of $\omega_1$(1GHz) and $\omega_2$(1.04GHz) are simultaneously applied to device Pt-S1. (**f**) Schematic of the wireless microwave detection experiment setup, involving a pair of antennas for signal transmission and reception and a voltmeter connected transversely. (**g**) Measured transverse second-harmonic output power vs input source power at room temperature. The solid line represents the quadratic relationship between output and input power.